\documentclass[showpacs]{article}
\usepackage{epsfig,epsf,graphics,color,rotate,amsmath}
\usepackage[latin1]{inputenc}
\usepackage[T1]{fontenc}

\begin{document}

\title{Experimental evidence of Alfv\'en wave propagation in a Gallium alloy}
\date{\today}
\author{Thierry Alboussi\`ere$^{1,2}$, Philippe Cardin$^1$, Fran\c cois Debray$^3$, Patrick La Rizza$^1$, \\ Jean-Paul Masson$^1$, Franck Plunian$^1$, Adolfo Ribeiro$^4$, Denys Schmitt$^1$ \\[5mm]
{\small $^1$ ISTerre, Institut des Sciences de la Terre, CNRS, Observatoire de Grenoble,} \\ {\small Universit\'e Joseph Fourier, Maison des G\'eosciences, BP 53, 38041 Grenoble Cedex 9, France } \\[3mm]
{\small $^2$ LGL, Laboratoire de G\'eologie de Lyon, CNRS, Universit\'e Lyon 1, ENS-Lyon,} \\ {\small G\'eode, 2 rue Rapha\"el Dubois, 69622 Villeurbanne, France} \\[3mm]
{\small $^3$ LNCMI, UPR 3228, CNRS-UJF-UPS-INSA, 25 rue des Martyrs, 38042 Grenoble, France} \\[3mm]
{\small $^4$ LIMSI, CNRS, BP 133, 91403 Orsay Cedex, France}
}

\maketitle

\begin{abstract}
Experiments with a liquid metal alloy, galinstan, are reported and show clear evidence of Alfv\'en wave propagation as well as resonance of Alfv\'en modes. Galinstan is liquid at room temperature, and although its electrical conductivity is not as large as that of liquid sodium or NaK, it has still been possible to study Alfv\'en waves, thanks to the use of intense magnetic fields, up to 13~teslas. The maximal values of Lundquist number, around 60, are similar to that of the reference experimental study by Jameson \cite{jameson}. The generation mechanism for Alfv\'en waves and their reflection is studied carefully. Numerical simulations have been performed and have been able to reproduce the experimental results despite the fact that the simulated magnetic Prandtl number was much larger than that of galinstan. An originality of the present study is that a poloidal disturbance (magnetic and velocity fields) is generated, allowing us to track its propagation from outside the conducting domain, hence without interfering. 
\end{abstract}

\section{Introduction}
\label{intro}

Alfv\'en waves are 
velocity and magnetic waves which propagate in electrically
conducting fluid along magnetic field lines. After their 
 discovery by H. Alfv\'en \cite{alfven42} in 1942 as a theoretical possibility from Maxwell's and Navier-stokes equations, they have been first ignored for a few years and then universally accepted as 
 a key ingredient for transporting energy and 
momentum in many astrophysical and geophysical fluid 
systems. Alfv\'en waves have since been observed in the magnetosphere of the Earth \cite{Tsurutani05}, in the solar wind, in the solar corona, in interplanetary
plasmas. These
oscillations are generally coupled to other physical
phenomena, generate non linear interactions and turbulence
which make their understanding and even their observation
 quite difficult \cite{VCGPC08}.   

Planetary cores are made of liquid metals and most of
them are dynamos \cite{stevenson03}. They generate strong magnetic
fields where Alfv\'en waves may propagate  \cite{hide66,braginsky75}.  
 Under the rotational constraint, the Alfv\'en waves
in rotating planetary core may take a degenerate form,
known as torsional waves and other quasi-geostrophic waves, and may be responsible for the
secular variations of the geomagnetic field \cite{canet09,gillet10}.
Moreover, small-scale turbulence in planetary cores is likely to be affected by Alfv\'en waves and this has implications on the rate of energy dissipation \cite{plunian2010}.  

Despite their astrophysical and geophysical importance, Alfv\'en waves have not been studied extensively in the laboratory. 
In plasmas, it is a necessary condition to have a collisionless plasma, hence Alfv\'en wave frequency smaller than the ion cyclotron frequency, to avoid excessive collision damping. A consequence of Alfv\'en wave dispersion relation -- with wave velocity independent of wavenumber modulus -- is that large wavelength must be produced and observed which necessitates specific devices such as the LAPD \cite{Carter06}.   
A review of early experiments on plasma Alfv\'en waves has been made by Gekelman \cite{gekelman99}.  
In liquid metals, the difficulty arises from Joule dissipation of magnetic energy. It can be escaped only with large dimensions and large applied magnetic fields. Waves have been identified in a spherical Couette flow with liquid sodium and a imposed dipolar magnetic field \cite{dtsjfm} and are thought to be magneto-inertial waves. 
Surprisingly, Alfv\'en waves have also been studied in solid state physics. More precisely, they have been observed in the cold plasma (holes and electrons) within low-temperature single crystals of bismuth \cite{HessHinsch73}.  

   Experimental detection in liquid metals have been 
undertaken in Stockholm after their discovery, first in mercury
 \cite{lundquist} and then in sodium \cite{lehnert53} but rather limited effects
 have been measured because of heavy damping. In
a nicely designed sodium experiment, in a torus, 
Jameson \cite{jameson} has been able to produce strong resonant effects
at the fundamental frequency of the Alfv\'en mode and a
weak resonance at three times this frequency in impressive agreement
with his theoretical prediction, both in terms of frequency and amplitude of resonance. However, his results on propagation of Alfv\'en waves have not been published, except in his PhD thesis \cite{JamesonPhD}.

Now, high magnetic fields have become available for industrial and experimental purposes and we take this opportunity to perform experiments in a small setup, using a gallium alloy at room temperature and yet achieving a value for the Lundquist number comparable to that of Jameson. This is a first step in anticipation of future liquid sodium experiments where the Lundquist number might be increased by a factor of order 10. 

After an introduction to the properties of theoretical Alfv\'en waves (section \ref{theorie}), we present the experimental results. 
In our design (section \ref{configuration}), an Alfv\'en wave is initiated by a short impulse of electrical current in a coil (Ec) placed just next to volume of liquid metal. The Alfv\'en wave is then observed thanks to the associated change in magnetic flux it generates through some ``passive'' coil (Tc), in which the electromotive force (EMF) is measured and recorded via a data acquisition system (sections \ref{rawpulse} and \ref{follow}). The observations are compared to numerical calculations. In addition, the response to a harmonic current is recorded (sections \ref{rawharmony} and \ref{reconstruction}) and also confronted to numerical results. The numerical schemes used are presented in section \ref{numpulse} and \ref{numharmony}. In a last section (\ref{discussion}), the results are discussed and some numerical calculations at significantly larger Lundquist number are shown. The possibility and expected advantages to upgrade the setup so that liquid sodium can be used are envisaged.    

\section{Theoretical Alfv\'en waves}
\label{theorie}

In a background magnetic field, ${\bf B}$, an electroconducting material can sustain electromagnetic waves. Let us consider a small magnetic disturbance ${\bf b}$ associated to a small velocity field ${\bf u}$. In a uniform imposed magnetic field, the linearized Navier-Stokes and induction equations can be written:
\begin{eqnarray}
\rho \frac{\partial {\bf u}}{\partial t} = - {\bf \nabla } p + \left( \frac{{\bf B}}{ \mu } \cdot {\bf \nabla}  \right) {\bf b} + \rho \nu {\bf \nabla }^2 {\bf u},  \label{nsdim} \\
\frac{\partial {\bf b}}{\partial t} = \left( {\bf B} \cdot {\bf \nabla}  \right) {\bf u} + \frac{1}{\mu \sigma} {\bf \nabla }^2 {\bf b}. \label{inductiondim} 
\end{eqnarray}
Linear combinaisons of these equations lead to:
\begin{eqnarray}
\frac{\partial {\bf u}^+}{\partial t} = - {\bf \nabla } \frac{p}{\rho} + \left( \frac{{\bf B}}{\sqrt{\rho \mu  }} \cdot {\bf \nabla}  \right) {\bf u}^+ + \nu {\bf \nabla }^2 {\bf u} + \frac{1}{\mu \sigma} {\bf \nabla }^2 \frac{{\bf b}}{\sqrt{\rho \mu}},  \label{uplus} \\
\frac{\partial {\bf u}^-}{\partial t} = - {\bf \nabla } \frac{p}{\rho} - \left( \frac{{\bf B}}{\sqrt{\rho \mu  }} \cdot {\bf \nabla}  \right) {\bf u}^- + \nu {\bf \nabla }^2 {\bf u} - \frac{1}{\mu \sigma} {\bf \nabla }^2 \frac{{\bf b}}{\sqrt{\rho \mu}},  \label{umoins} 
\end{eqnarray}
where ${\bf u}^\pm = {\bf u } \pm {\bf b}/\sqrt{\rho \mu}$ are the so-called Elsasser variables. One can get rid of the pressure term by taking the curl of the equations, but this term is also identically zero for the case of simple transverse shear structures. Furthermore, in the limit ideal case of vanishing diffusivities (kinematic viscosity and magnetic diffusivity), one obtains pure wave equations without damping
\begin{equation}
\frac{\partial {\bf u}^\pm}{\partial t} = \pm \left( \frac{{\bf B}}{\sqrt{\rho \mu  }} \cdot {\bf \nabla}  \right) {\bf u}^\pm, \label{pureAlfven}
\end{equation}
The Elsasser variable ${\bf u}^+$ propagates in the direction opposite to ${\bf B}$ while ${\bf u}^-$ propagates in the direction of ${\bf B}$. The wave velocity (phase velocity) is the Alfv\'en velocity ${\bf B}/\sqrt{\rho \mu}$ and the structure of the wave equation is so simple that its group velocity is also equal to the Alfv\'en velocity, independently of the direction and magnitude of the wave vector. In a '${\bf u}^+$' wave-packet, ${\bf u}^-={\bf 0}$ (if not, it would have split into two wave packets of opposite velocity), and conversely, so that equipartition of kinetic and magnetic energy is established. 

In the case of liquid metals (and some plasmas), magnetic diffusivity is much larger than kinematic viscosity so that a diffisivity term $1/(\mu \sigma ) {\bf \nabla}^2 ({\bf u}^+ - {\bf u}^-)$ is added to the '${\bf u}^+$' equation and $1/(\mu \sigma ) {\bf 
\nabla}^2 ({\bf u}^- - {\bf u}^+)$ to the '${\bf u}^-$' equation. The ratio of orders of magnitude of the advective term $({\bf B}/\sqrt{\rho \mu} \cdot {\bf \nabla}) {\bf u}^\pm $ to the diffusive term, is the Lundquist number 
\begin{equation}
Lu = \sqrt{\frac{\mu }{ \rho}} \sigma B L, \label{lundquist}
\end{equation}
for a given length-scale $L$. For a large Lunquist number, a wave packet of dimension $L$ can travel a distance of order $Lu \, L$ before it is dissipated by ohmic losses (within one magnetic diffusion time-scale $\mu \sigma L^2$).

\section{Experimental ``Galalfv\'en'' set-up}
\label{configuration}

\begin{figure}
\hspace*{3 cm}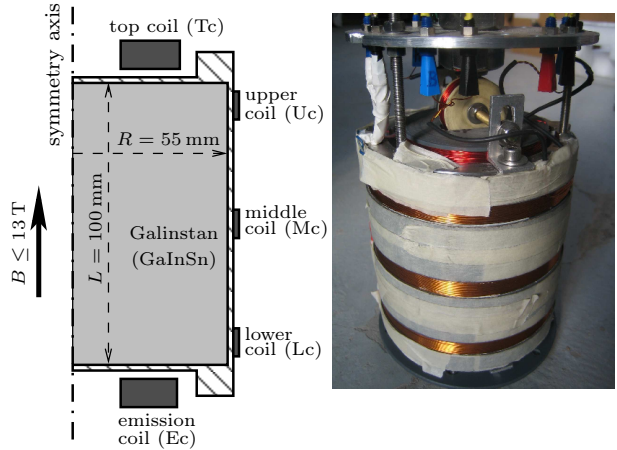 
\caption{Schematic drawing of the liquid-metal vessel and a photograph of the set-up}
\label{sketch}
\end{figure}

The setup consists in a cylindrical
tank of diameter 110~mm and of height $L$~=~100mm
as shown in Fig.~\ref{sketch}. The cavity is filled with Galinstan (eutectic 
 ternary alloy made of Gallium 68~\%, Indium 21~\% and Tin 
12~\%). Its melting point is 10.5 $^\circ$C which explains
the use of Galinstan, liquid at room temperature, instead of gallium (melting point 29.8 $^\circ$C). Its density is
$ \rho = 6370$~kg/m$^3$, its kinematic viscosity $\nu = 3.7 \, \times \, 10^{-7}$m$^2$s$^{-1}$ and its electrical conductivity $\sigma$ varies in
the literature between 2.3$\, \times \, 10^6$ \cite{LHLBLDLG06} and 3.46$\, \times \, 10^6$~$\Omega ^{-1}$m$^{-1}$ \cite{KKCS06}. In this work, we choose to take $\sigma = 3.35\, \times \, 10^6 $~$\Omega ^{-1}$m$^{-1}$ \cite{BuhlerMuller,Lyon} at 15~$^\circ$C.
The container is made of non-magnetic 316 stainless steel 
 of electrical conductivity  1.35$\, \times \, 10^6$~$\Omega ^{-1}$m$^{-1}$.
The wall thickness is 2~mm. During an experiment, the
box is placed in the bore (130~mm of diameter) of a powerful 
 (12~MW ) resistive electromagnet (magnet M5), facility of the
LNCMI in Grenoble \cite{GHMFL}. The electromagnet produces a
magnetic field of intensity between 0 and 13~T aligned with
the axis of the cylindrical container. Variations of the 
intensity of the imposed magnetic field along the axis of the
bore are less than 0.7 \% \cite{GHMFL} within the volume of the fluid.

A small magnetic disturbance is generated by a copper coil, the 'excitation coil' (Ec), mechanically attached under the container, in addition to the stationary imposed magnetic field.  This excitation coil, concentric with the container, 
consists of 300 turns of wires between 45 and 85~mm of diameter
and 10~mm of thickness.
Two types of experiments have been run, depending on the type of signal generated in the excitation coil: "pulse" experiments and "sweep" experiments.   
   To study the propagation of a magnetic perturbation ("pulse")
 we imposed a voltage to the coil
such that one obtains a one-signed impulse of electrical current in the
coil. The current is generated by a function generator (Agilent 33220 A) coupled to a linear amplifier restricted to 25~V (peak to peak) and 3~A.  A positive voltage (20~V) 
 is applied to the coil, followed by a slightly shorter negative voltage. 
The total duration is 300~$\mu$s. 
It produces an impulse of electrical current as shown in
Fig.~\ref{pulsesdim} of 0.25~A (peak value) which is an approximation for a Dirac function, as long as it is short compared to the duration of Alfv\'en wave propagation. 
This is a necessary condition for the response not to depend on the exact shape of the pulse function, {\it i.e.} there is no need for a dimensionless parameter characterizing the duration of the pulse compared to that of Alfv\'en wave propagation in addition to the Lundquist number. This condition is well fulfilled for imposed magnetic fields of a few teslas but the duration of the pulse is more than a third of the time for Alfv\'en waves to reach the other end of the cavity when the magnetic field intensity is maximum (13~T). 
This electrical current "pulse" generates a poloidal magnetic field
at the bottom of the container of order of 1~mT . Each run is
made of 200 pulses of 300~$\mu$s shot every 101.2~ms so that they are all independent from each other. The 200
signals are stacked to reduce the noise and filtered around
50~Hz and its harmonics to eliminate the frequency of the
electrical network. 
   A "sweep" run is designed to measure the resonance of Alfv\'en 
modes in the cavity by imposing a sinusoidal current to the
excitation coil with a frequency varying exponentially 
 in time from 20~Hz to 4000~Hz over a duration of 10~s. 

The excitation current first generates a magnetic field within the liquid gallium alloy which then propagates as an Alfv\'en wave towards the opposite end of the cylinder. 
The signals acquired are global measurements of magnetic flux variation through four different axisymmetrical coils (see Fig.~\ref{sketch}): this signal is multiplied by the number of turns in each coil. The so-called lateral coils, Lower Coil (Lc), middle coil (Mc) and upper coil (Uc) have 30 turns and a diameter just over 110~mm, and their mean axial position is 8~mm, 50~mm and 92~mm respectively, measured from the bottom of the fluid cavity. The coil at the top (Tc) is identical to the excitation coil (Ec) with 300 turns, a mean radius of 65~mm and an average axial position 10~mm above the fluid cavity. Those four voltages and the current in the excitation coil (measured via a calibrated shunt) were recorded using a NI A/D data acquisition card (16 bit resolution) monitored by a Labview programme installed on a pentium PC. The rate of acquisition was set to 20~kHz. 

\section{Magnetic pulses: experimental results of propagation}
\label{rawpulse}

Let us first describe with hand-waving arguments the sequence of events after a pulse of current is generated in the excitation coil, as revealed by the analysis of experimental results and the visualization of numerical results. In the excitation coil, the electric current increases to a maximum and then decreases to zero (neglecting that short phase where it becomes slightly negative). While the current is increasing, its associated magnetic field penetrates into the gallium alloy by magnetic diffusion (the timescale is shorter than Alfv\'en wave propagation). Maxwell equation ${\bf \nabla} \times {\bf E} = -\partial {\bf B} / \partial t$ indicates that an electric field is generated, which forces an electric current to flow in the azimuthal direction and opposite to the direction of the excitation current. Alfv\'enic propagation makes it move immediately. During the subsequent phase of decreasing electric current, another electric current loop is generated below the initial one with the same direction as the excitation current. This state is thus the initial state for Alfvenic propagation: two current loops in opposite azimuthal directions, one below the other one. Their associated poloidal magnetic field consists in a double torus of opposite poloidal fields. 

\begin{figure}
\vspace*{-3 cm}
\hspace*{3 cm}\includegraphics[width=10 cm, keepaspectratio]{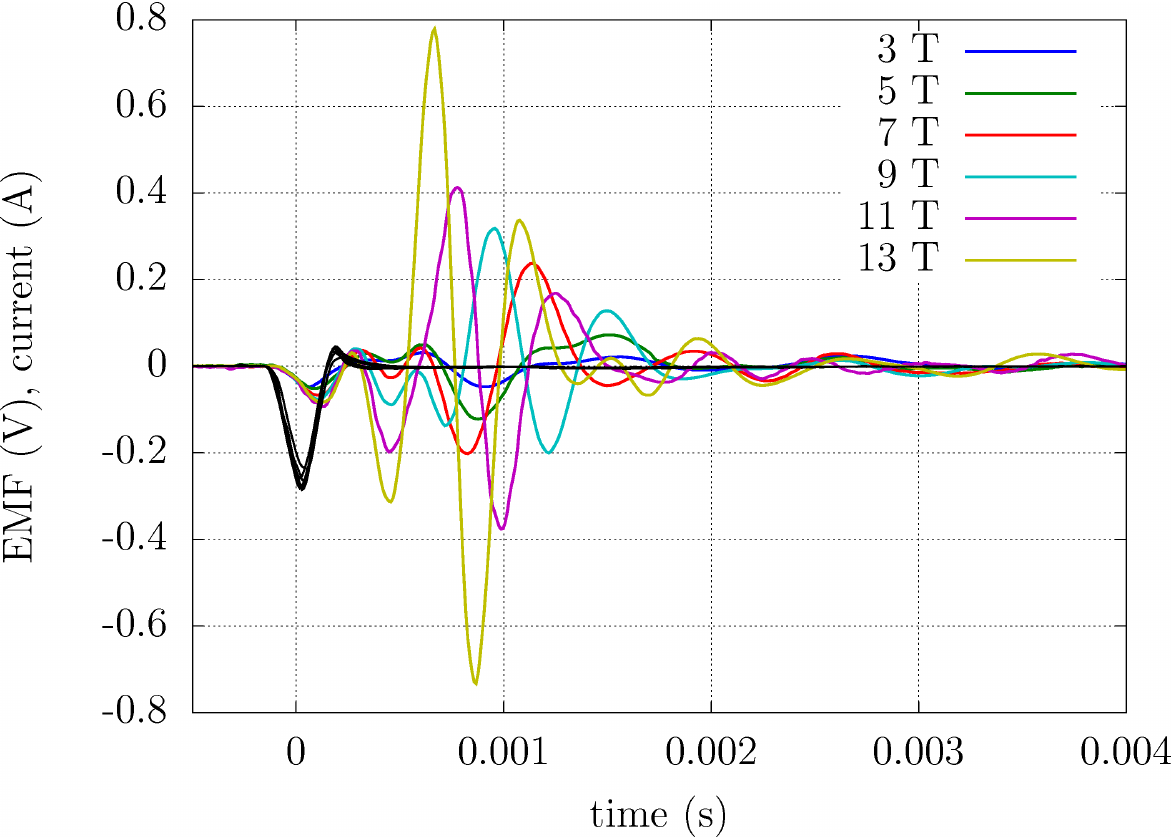}
\vspace*{-0 cm}
\caption{A pulse of electrical current (black curve) in the emitting coil generates an Alfv\'en wave. The associated change in magnetic flux is recorded at the opposite end in the top coil, Tc}
\label{pulsesdim}
\end{figure}

Figure \ref{pulsesdim} shows the electric potential recorded at the top coil (Tc) following a current pulse in the Emission coil (Ec) for different values of the applied magnetic field (B = 3, 5, 7, 9, 11, 13 teslas). It can be seen that the maximal amplitude response is attained at shorter and shorter times when the background magnetic field intensity is increased. The amplitude of that response increases with increasing values of the magnetic field. These findings are compatible with the propagation of an Alfv\'en wave of velocity proportional to the applied magnetic field, and subjected to Joule dissipation. 
They 
propagate along the applied magnetic field. Anticipating on section 
\ref{numpulseAdolfo}, snapshots at different times are represented in figure \ref{reflect}.

\begin{figure}
\vspace*{-3 cm}
\hspace*{-0.5 cm}\includegraphics[width=8 cm, keepaspectratio]{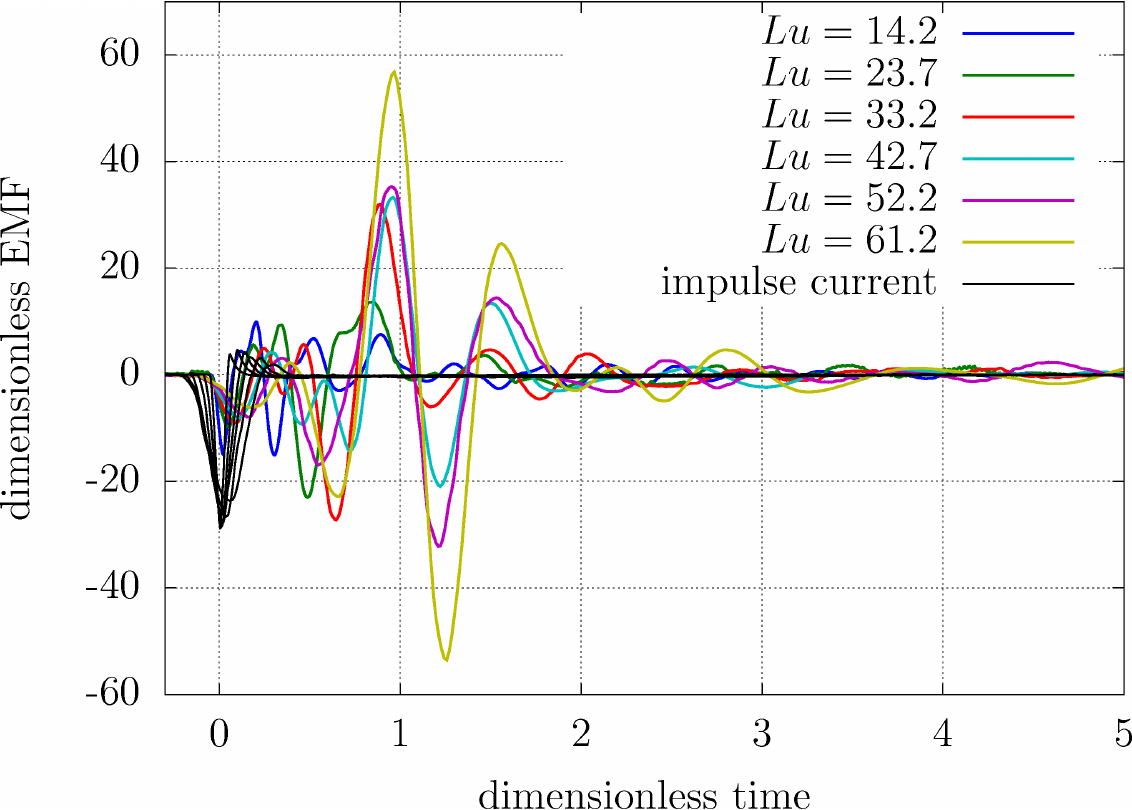} \includegraphics[width=8 cm, keepaspectratio]{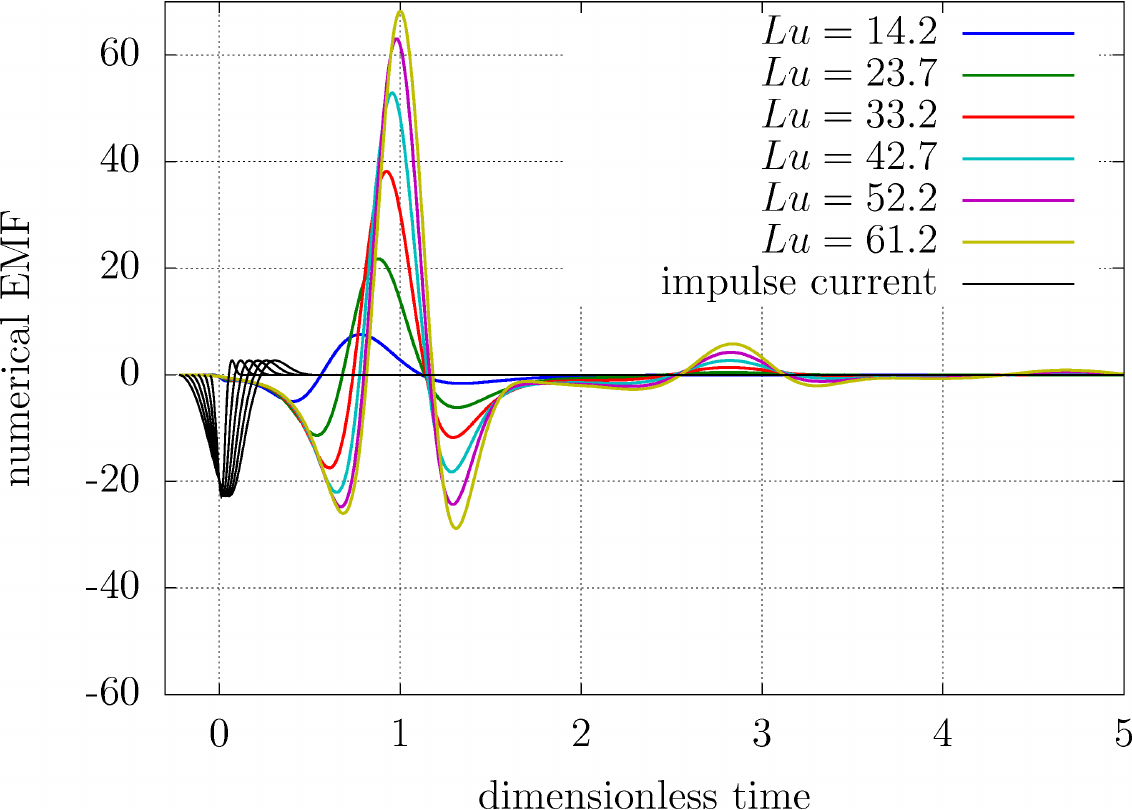}
\vspace*{-0 cm}
\caption{Same as Fig.~\ref{pulsesdim} on the left hand side, except time has been made dimensionless using the propagation time of Alfv\'en waves and EMF has been made dimensionless using $Lu \, ( n\, I)  / (\sigma \, L )$. On the right hand side, the same pulses computed numerically}
\label{pulsesadim}
\end{figure}

From dimensional analysis (Buckingham's theorem), the problem of wave propagation must involve three independent dimensionless numbers, as seven dimensional scales are sufficient to define the problem: $L$, $B$, $\mu$, $\sigma$, $\rho$, $\nu$ and $nI$, the length-scale, imposed magnetic field, magnetic permeability, electrical conductivity, kinematic viscosity and scale for the intensity of the current generated in the excitation coil ($n$ turns of intensity $I$) respectively, while four fundamental units are needed: meter, kilogram, second and Ampere. We are free to choose those three dimensionless numbers. A good choice is $P_m=\mu \sigma \nu $, $Lu = \sqrt{{\mu}/{\rho}} \sigma B L$ and $A=\mu n I/( B L)$, the magnetic Prandtl number, Lundquist number and a dimensionless number characterizing the relative intensity of the magnetic field associated to the pulse to that of the imposed magnetic field. For galinstan, the magnetic Prandtl number is $P_m = 1.56 \, \times \, 10^{-6} $. Its small value implies that viscous dissipation is negligible compared to Joule dissipation and that Alfv\'en wave propagation will not be affected by viscous effects. The number $A$ is also irrelevant (or rather, the experiments are done in the limit of weak forcing) since the available regime of Alfv\'en waves has been checked to be strictly linear. When the intensity of the current pulse is changed, all response signals scale exactly with the intensity of the pulse. Hence, the Lundquist number is the only relevant dimensionless number governing Alfv\'en wave propagation in liquid metals.

The same results as in Fig.~\ref{pulsesdim} are plotted in Fig.~\ref{pulsesadim} where the x-coordinate has been changed for a dimensionless measure of time, based on the flight-time of Alfv\'en waves $L / V_A$ and where the legend is expressed in terms of dimensionless Lundquist numbers instead of magnetic field intensity. In addition, the EMF has been made dimensionless using a scale for the magnetic flux $\mu \, n\, I \, L$ divided by the Alfv\'en flight-time, {\it i.e.} $\sqrt{\mu / \rho } B \, n\, I \ = \ Lu \, n\, I / (\sigma \, L)$ . The phase velocity of Alfv\'en waves in the axial direction is $V_A = B / \sqrt{\rho \mu}$, where $\mu = 4 \pi 10^{-7}$H$\,$m$^{-1}$ is the permeability of vacuum (suitable for the non-magnetic galinstan alloy) and $\rho = 6370\, $kg$\,$m$^{-3}$ is the density of galinstan at room temperature. With B~=~13~T and L~=~10~cm, Alfv\'en speed is $V_A \simeq 144.5\, $m$\,$s$^{-1}$ and the theoretical flight-time along the length of the cylinder is $L / V_A \simeq 0.692\, $ms, in accordance to the observed time of response in Fig.~\ref{pulsesdim}. In dimensionless time (Fig.~\ref{pulsesadim}), it is seen that the reception signal at the end of the cylinder has its maximum response around dimensionless time unity and that the signals for different values of the magnetic field are similar except for an increase of amplitude with increasing strength of the applied magnetic field. There is little evidence of a signal at dimensionless time 3, when the wave should reappear after 2 reflections, except for B~=~13~T, due to excessive damping.    

\section{Harmonic response: experimental results}
\label{rawharmony}

The harmonic response could have been obtained step by step by imposing a sine function current to the excitation coil with each possible frequency. Instead, we chose to apply a so-called "sweep" function, which is a sine function with slowly evolving frequency from a minimum of 20~Hz to a maximum of 4.0~kHz. We have made use of an option of the function generator ("log" variation) which is to have the frequency increase exponentially in time: this means that the change in frequency is always the same during each period of the signal, in our case $0.52\,$Hz. This is satisfying, even at the lowest frequency 20~Hz, since the relative change in frequency is only 25~\% after ten periods of the signal. The higher the frequency, the smaller is the relative change. One can safely assume that the response to the sweep is close to that of a large set of purely harmonic responses: this has been checked for several values of frequency. After data processing to determine frequency and response amplitudes, the available range of frequencies is reduced to 30--3500~Hz. 

\begin{figure}
\vspace*{-3 cm}
\hspace*{3 cm} \includegraphics[width=10 cm, keepaspectratio]{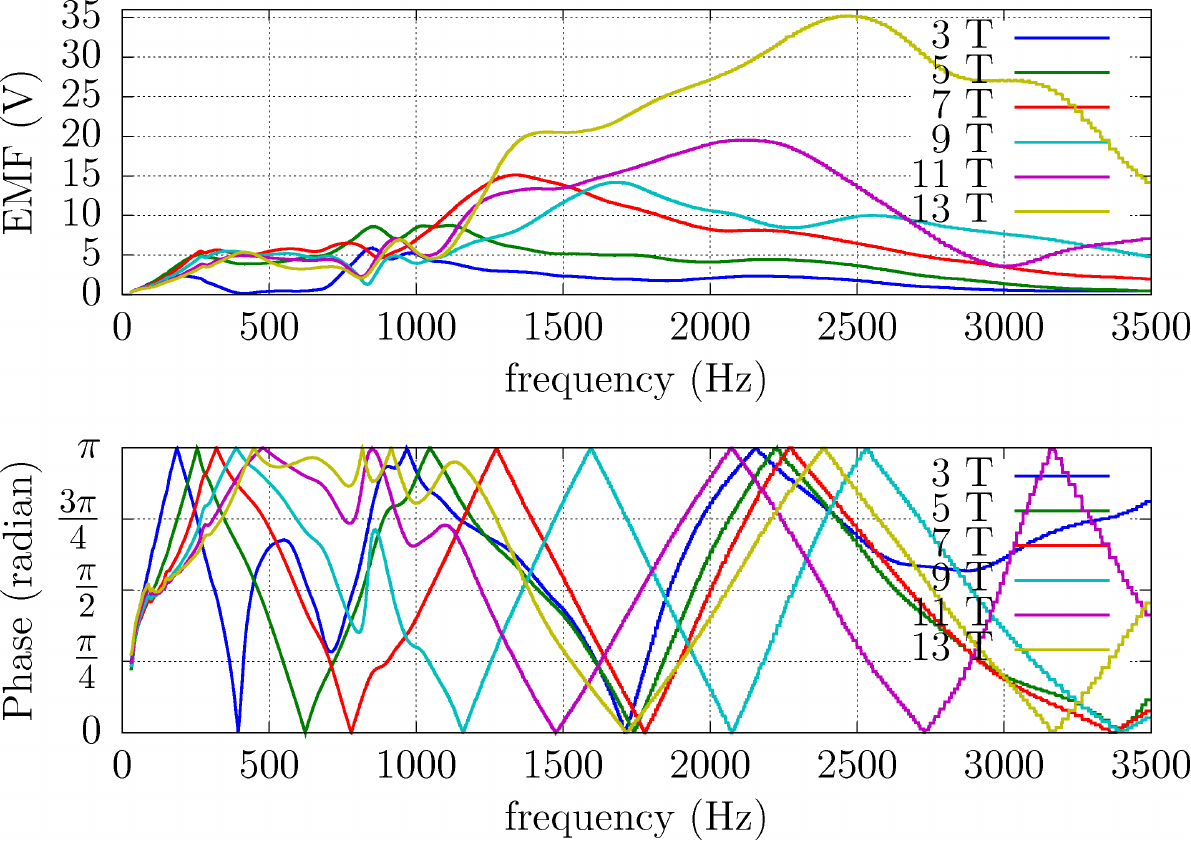}
\vspace*{-0.0 cm}
\caption{Response to a sweep in the emitting coil, measured at the top coil (Tc)}
\label{sweepdim}
\end{figure}

For each coil, the response to the sweep consists in an amplitude and in a phase shift for each frequency between 30 and 3500~Hz. 
In fig.~\ref{sweepdim}, both contributions to the response to the sweep are shown for the top coil (Tc), for the same set of magnetic field values as for the pulses: 3, 5, 7, 9, 11 and 13~T. 
The response moves towards higher frequencies for stronger applied magnetic fields, while getting bigger.

\begin{figure}
\vspace*{-0.0 cm}
\hspace*{-0.5 cm}\includegraphics[width=8 cm, keepaspectratio]{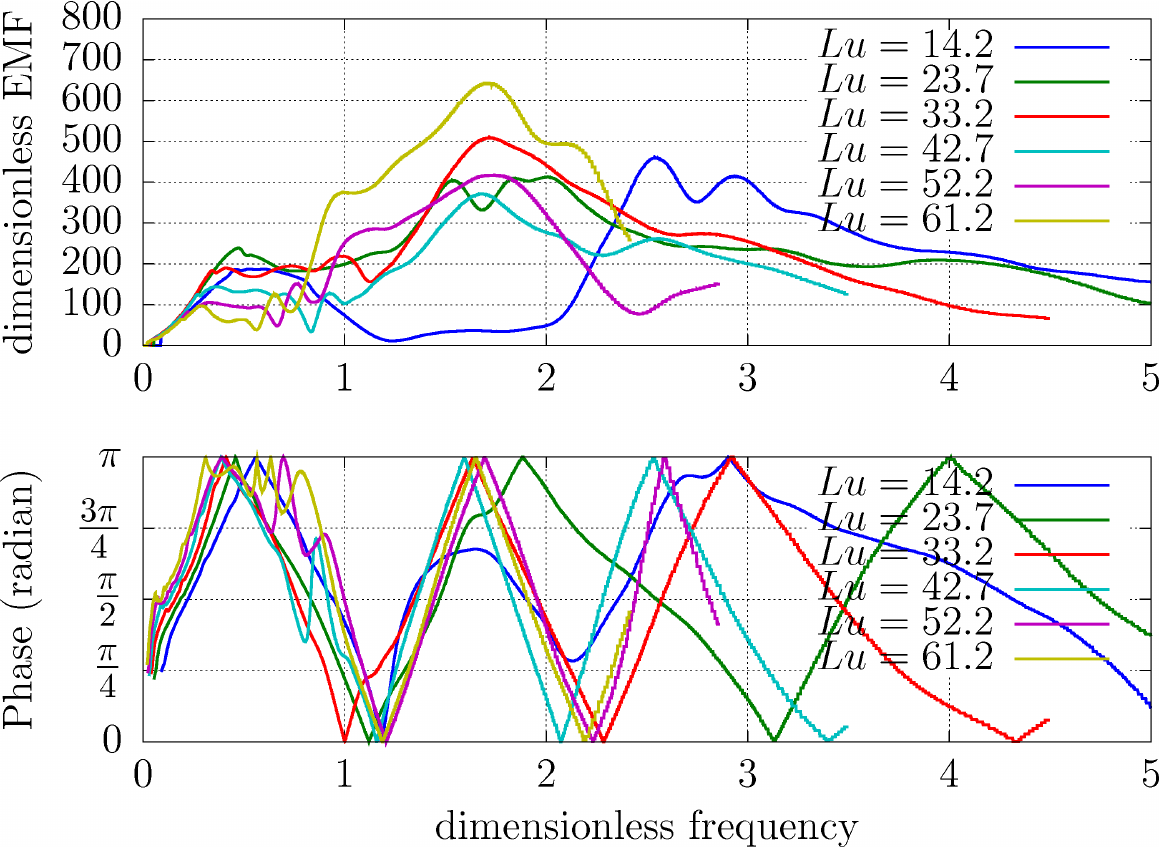} \hspace{5 mm} \includegraphics[width=8 cm, keepaspectratio]{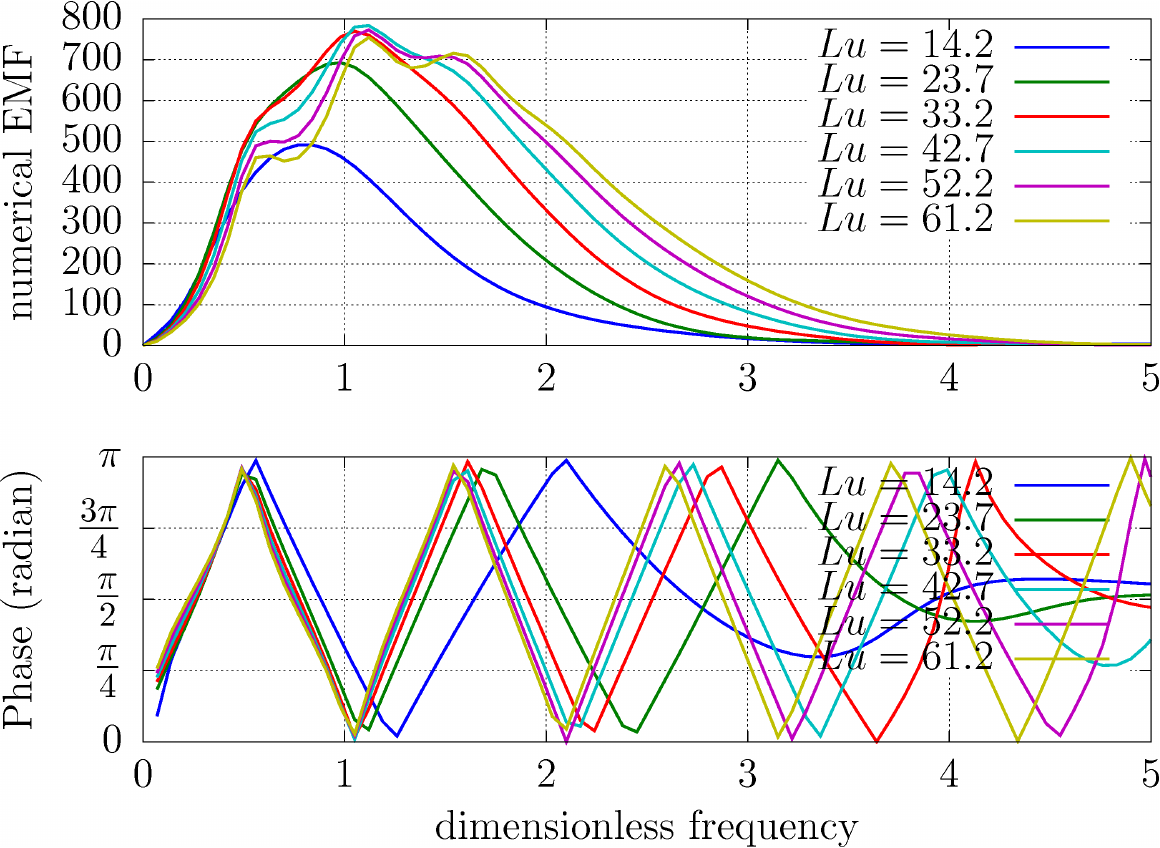}
\vspace*{-0 cm}
\caption{Same as Fig.~\ref{sweepdim} on the left hand side, with dimensionless frequency and EMF. On the right hand side, numerical simulations of harmonic responses.}
\label{sweepadim}
\end{figure}

The same data as in Fig.~\ref{sweepdim} are plotted in Fig.~\ref{sweepadim}, except that frequencies are now made dimensionless using the inverse of the Alfv\'en flight-time, {\it i.e.} $f_A=V_A / L$, which is around $1445$~Hz for B~=~13~T. Also, the amplitude of the meassured EMF has been divided by the magnitude $B$ of the imposed magnetic field: assuming that the intensity of the magnetic signal carried by the wave is independent of $B$, and given that the propagation speed of this signal is proportional to $B$ (Alfv\'en velocity), it is expected that the EMF is proportional to $B$, as it corresponds to the time derivative of the magnetic flux. A good collapse of the response is found for large applied magnetic fields. The curve corresponding to the smallest Lundquist number $Lu=14.2$ is clearly distinct from the others. We do not expect to see clearly an Alfv\'enic response at such a low Lundquist number. Moreover vibration disturbances have a large relative contribution at low Lundquist number and are probably responsible for this large departure. 

The phase shift is particularly simple and increases linearly in time, as expected for waves with uniform velocity irrespective of the wavenumber. The sign of the slope changes at $0$ and $\pi$ because we have restricted the phase shift to be within the interval $[0; \pi ]$. the phase shift seems to be less affected by disturbances (like vibrations) than the amplitude of the response. This might be due to the fact that disturbances have a random phase shift and no clear effect on the outcome.

When the imposed magnetic field is strengthened, diffusion effects become comparatively weaker. It is thus not expected that the responses for different values of the magnetic field will collapse exactly on Fig.~\ref{sweepadim}. In addition, mechanical vibrations of the experiments induce some unwanted contributions on the measured signals. These vibrations are partly deterministic as they are triggered by the forces associated to the electric current pulse. However, their subsequent evolution is independent of the applied magnetic field and their associated signal is due to mechanical displacement of the setup within the imposed magnetic field. 

\section{Pulse reconstruction from harmonic response}
\label{reconstruction}

\begin{figure}
\vspace*{-2 cm}
\hspace*{3 cm}\includegraphics[width=10 cm, keepaspectratio]{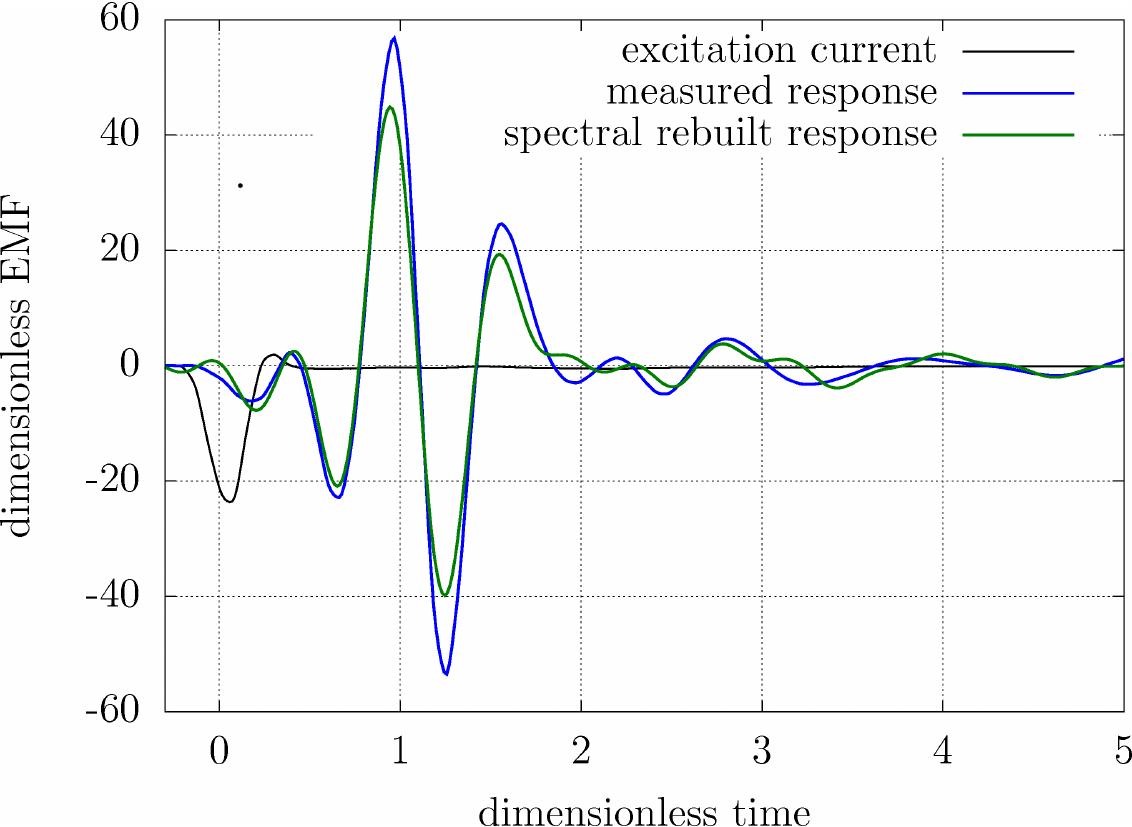}
\vspace*{-0 cm}
\caption{A synthetic response to the pulse in the coil is calculated from the harmonic response and plotted for comparison next to the actual response to the pulse ($B=13$~T)}
\label{reconstFig}
\end{figure}

As discussed above, the amplitude of the magnetic disturbance that is created by the excitation coil is so small compared to the imposed uniform magnetic field of the magnet that our experiments are in a linear regime. This has been checked indeed and the measured signals vary linearly with the amplitude of the imposed current with no detectable departure from linearity. 

A consequence of linearity is that our pulse and sweep experiments are not independent from each other. In fact, the harmonic response from the sweep experiments contains the whole information regarding this linear system and the pulse response can be reconstructed from it. Let ${\cal{C}}(t)$ be the current in the excitation coil as a function of time during a pulse experiment. Its Fourier transform can be written:
\begin{equation}
{\cal{C}}^{FT} (f) = \int_{-\infty}^{\infty} {\cal{C}}(t) e^{i 2 \pi f t} dt \label{FT} 
\end{equation}
Note that, because ${\cal{C}}(t)$ is real, ${\cal{C}}^{FT} (f)$ must obey the conjugate symmetry ${\cal{C}}^{FT} (-f) = \left({\cal{C}}^{FT} (f) \right) ^{*}$.
Next step, the Fourier transform of the current is multiplied by the complex harmonic response ${\cal{H}} (f)$ (the real part is the in-phase response, while the imaginary part is the out-of-phase response): 
\begin{equation}
{\cal{R}}^{FT} (f) = {\cal{H}} (f) \, {\cal{C}}^{FT} (f) \label{RFT}
\end{equation}
This is the Fourier transform of the actual response to the pulse ${\cal{C}}(t)$. It must then be conjugate symmetric, and as ${\cal{C}}^{FT} (f)$ obeys that symmetry so must also ${\cal{H}} (f)$. The last step is to take the inverse Fourier transform and get the actual response to the initial pulse:
\begin{equation}
{\cal{R}} (t) = \int_{-\infty}^{\infty} {\cal{R}}^{FT}(f) e^{- i 2 \pi f t} df  \label{IFTRFT}
\end{equation}
That signal is called the synthetic response to the pulse calculated from the harmonic response. In Fig.~\ref{reconstFig}, the synthetic signal is compared to the actual signal measured during a pulse experiment, for a magnetic field intensity of 13~T. The signals compare well but are different in a few ways. The synthetic signal fails to recover the maximal amplitude and there are small additional bumps. These differences can be explained by two arguments: first, the signals are discretized and secondly, the range of frequencies covered during the sweep experiments is not infinite. In particular, at the highest frequency (3500~Hz), the harmonic response is still large for strong magnetic fields. 
Ideally, we should have gone to higher frequencies.

\section{Following an Alfv\'en wave}
\label{follow}

\begin{figure}
\vspace*{-3 cm}
\hspace*{3 cm}\includegraphics[width=10 cm, keepaspectratio]{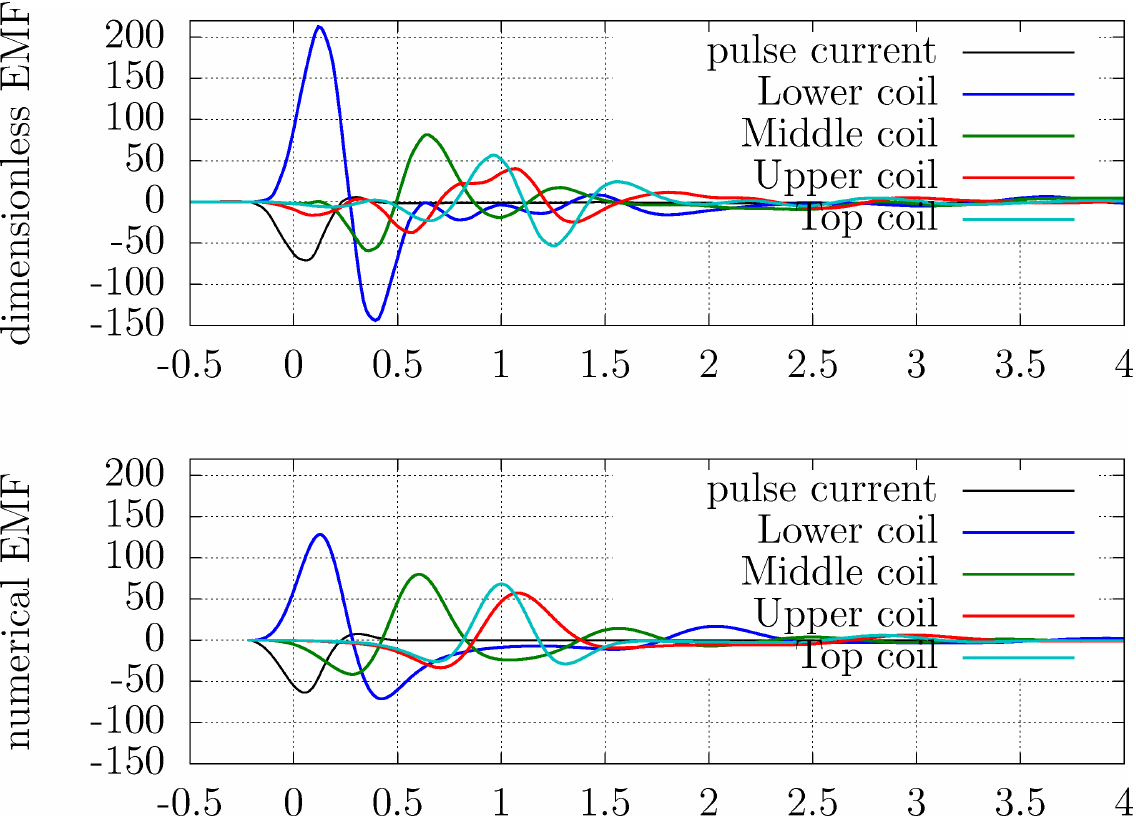}
\vspace*{-0 cm}
\caption{Electromotive force in the four coils Lc, Mc, Uc and Tc, following a pulse, at $B$~=~13~T, as a function of dimensionless time}
\label{followFig}
\end{figure}

With the three lateral coils and the top coil, it is possible to follow the progression of an Alfv\'en wave generated by a pulse. On Fig.~\ref{followFig} the four measured EMF (electromotive forces) are shown for the case $B$~=~13~T. 
Anticipating on section \ref{numpulse}, the signals are compared to numerical simulations.  
The linear response to a pulse is calculated using FreeFem, a 2D finite element software developed by INRIA \cite{freefem} and SFEMaNS \cite{Guer06}. 

The pulse signal seems to reach the top coil even before reaching the upper coil. The distance between each of those coils to the excitation coil is actually quite similar. In addition, the pulse is severely distorted during reflection (see Fig.~\ref{reflect} and \ref{diss}) which implies that the electromotive force is not necessarily similar for both coils.  

\section{Numerical simulations of pulse response}
\label{numpulse}

\begin{figure}
\vspace*{-6 cm}
\hspace*{3 cm}\includegraphics[width=10 cm, keepaspectratio]{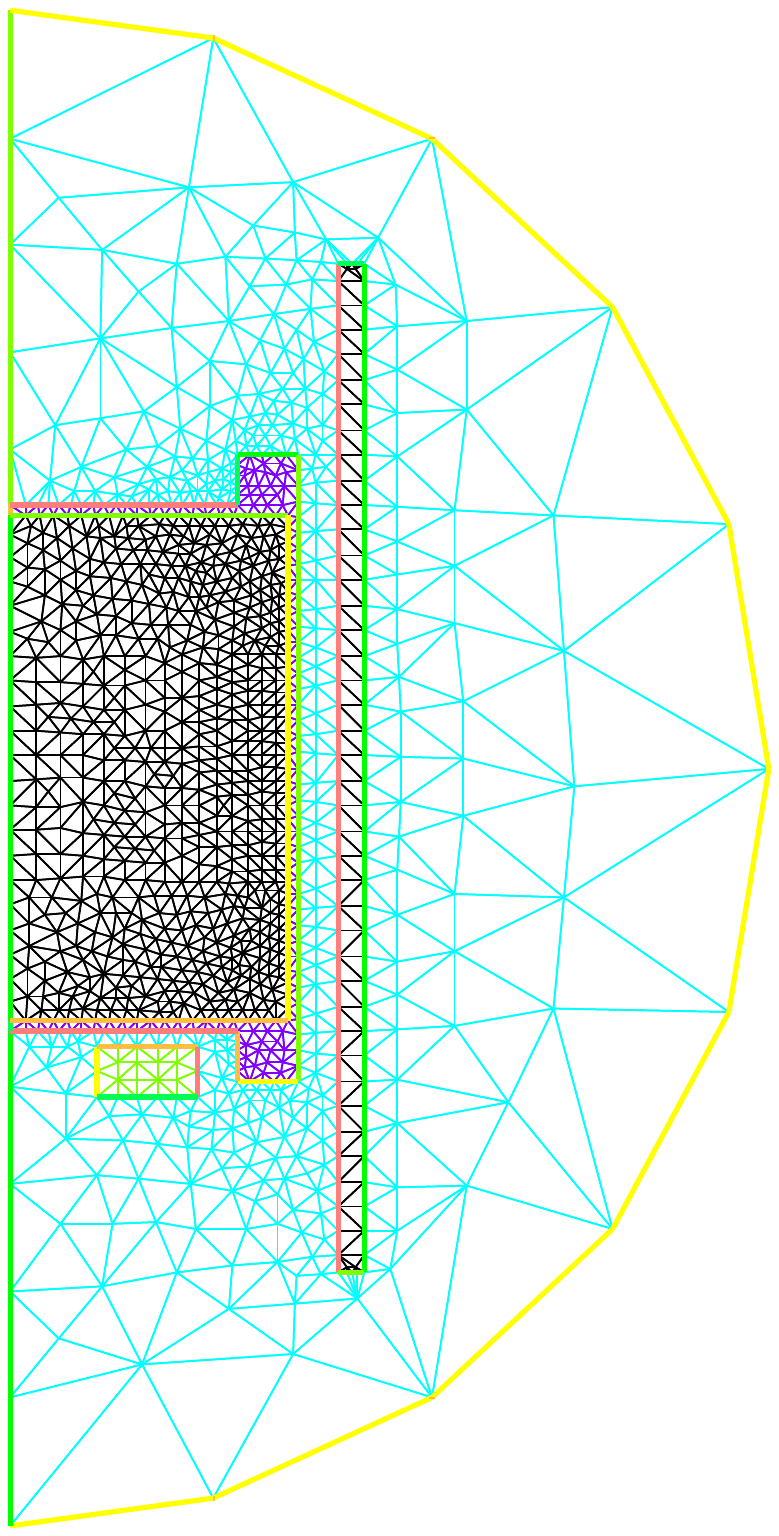}
\caption{Mesh used (coarse version for easy visualization) to calculate Alfv\'en waves with FreeFem++: meridional plane of an axisymmetrical geometry with symmetry axis on the left.}
\label{mesh}
\end{figure}

\begin{figure}[h]
\vspace*{-3 cm}
\hspace*{3 cm}\includegraphics[width=10 cm, keepaspectratio]{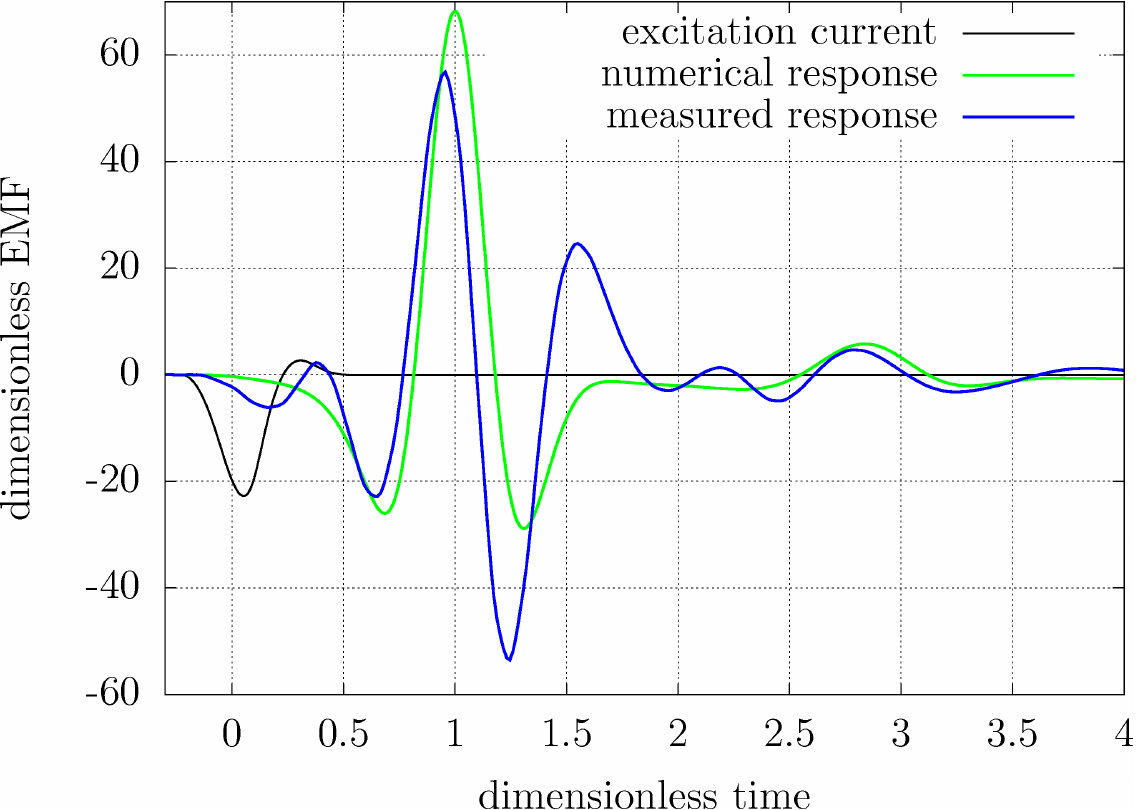}
\vspace*{-0 cm}
\caption{The numerically calculated pulse response is compared to the experimental measurements  ($B=13$~T)}
\label{comparenum}
\end{figure}

We have used two different sofwares to compute the response to electric current pulses, FreeFem++ and SFEMaNS. They are based on different numerical schemes and boundary conditions are not written using the same physical variables. Among the general boundary conditions are the magnetic conditions at the interfaces between the various domains of different electrical conductivity (galinstan, stainless steel, insulating regions). The meshes have been designed to achieve low magnetic Prandtl numbers, as thin Hartmann layers develop, and there is certainly a better control of the mesh grid with SFEMaNS. Conversely, magnetic and velocity continuity equations are maintained to small value using a penalty method in SFEMaNS, while they are exactly enforced in the FreeFem++ code. We felt reassuring that both numerical methods lead eventually to the same results. There exist numerical magnetohydrodynamic codes with uniform electrical conductivity, there also exist electromagnetic codes (calculation of high frequency inductors for material processing), but there are no fully-coupled magnetohydrodynamic codes capable to treat regions of different conductivities: this was an incentive to develop SFEMaNS which can also include domains of different magnetic permeabilities.  

\subsection{FreeFem++}
\label{freefemsub}

With FreeFem++, the numerical modelling was restricted to axisymmetric poloidal magnetic ${\bf b}$ and velocity ${\bf v}$  disturbances with reference to a basic motionless state with a strong uniform magnetic field ${\bf B}_0 = B_0 {\bf e}_z$ along the direction of the symmetry axis. In cylindrical coordinates $( \rho, \phi , z)$, the velocity and magnetic disturbances are written:
\begin{equation}
{\bf v} = \left[ \begin{matrix} v_\rho ( \rho , z) \\  0 \\ v_z ( \rho , z) \end{matrix}  \right] ,  \hspace{5mm}
{\bf b} = \left[ \begin{matrix} b_\rho ( \rho , z) \\  0 \\ b_z ( \rho , z) \end{matrix}  \right] , \label{velmag}  
\end{equation}
\begin{equation}
{\bf a} = \left[ \begin{matrix} 0 \\  a (\rho , z) \\ 0 \end{matrix}  \right]  , \hspace{5mm}
{\bf \omega} = {\bf \nabla} \times {\bf v} = \left[ \begin{matrix} 0 \\  \omega ( \rho , z) \\ 0 \end{matrix}  \right] . \label{aomega}
\end{equation}
 This analysis is restricted to a linearized approach, neglecting all quadratic terms involving ${\bf u}$ or ${\bf b}$, because the induced magnetic field ${\bf b}$ is small compared to $B_0$: this linearization is also supported by the experimental results. Using the length-scale $L$ and the Alfv\'en velocity based on $B_0$, we have a natural time-scale -- the Alfv\'en flight-time $t_A = L \sqrt{\rho \mu }/B_0 $. 
Dimensionless Navier-Stokes and induction equations are linearized as follows assuming a uniform and constant fluid density:
\begin{eqnarray}
\frac{\partial {\bf v}}{\partial t} &=& - {\bf \nabla } p + \left( {\bf \nabla }\times {\bf b} \right) \times {\bf e}_z + Lu^{-1} P_m {\bf \nabla } ^2 {\bf v} , \label{ns} \\
\frac{\partial {\bf b}}{\partial t} &=&  {\bf \nabla }\times \left( {\bf v} \times {\bf e}_z \right) + Lu^{-1} {\bf \nabla } ^2 {\bf b} . \label{induction}
\end{eqnarray}
The curl of linearized Navier-Stokes equation and the uncurled linearized induction equation can be written as follows in the azimuthal direction and in dimensionless variables:
\begin{eqnarray}
\frac{\partial \omega }{\partial t} &=& \frac{ \partial  j_\phi }{\partial z} + Lu^{-1} P_m \left[ \frac{\partial ^2 \omega }{\partial z^2} + \frac{\partial}{\partial \rho } \left( \frac{\partial \omega }{\partial \rho } + \frac{\omega}{\rho}  \right) \right] , \label{vort} \\
\frac{\partial a}{\partial t} &=& - v_\rho + Lu^{-1} \left[ \frac{\partial ^2 a}{\partial z^2} + \frac{\partial}{\partial \rho } \left( \frac{\partial a}{\partial \rho } + \frac{a}{\rho}  \right) \right] , \label{vectpot}
\end{eqnarray}
where $j_\phi$ the electric current density in the azimuthal direction can be expressed using Ohm's law with no azimuthal electric potential by symmetry:
\begin{equation}
j_\phi = - \frac{\partial a}{\partial t} - v_\rho . \label{Ohm}
\end{equation}
Velocity is expressed in terms of a streamfunction $\psi$, ${\bf v} = {\bf \nabla} \times \left( \psi (\rho, z) {\bf e}_\phi  \right)$. Its laplacian corresponds to the vorticity:
\begin{equation}
\omega = - \frac{\partial ^2 \psi }{\partial z^2 } - \frac{\partial }{\partial \rho } \left[ \frac{1}{\rho} \frac{\partial }{\partial \rho } \left( \rho \psi \right)  \right]. \label{psi}
\end{equation}

Equations (\ref{vort}), (\ref{vectpot}), (\ref{Ohm}) and (\ref{psi}) form a closed set equations for three unknown $\psi$, $\omega$ and $a$, with appropriate boundary conditions: no slip on the walls of the container and matching to the external harmonic magnetic field. These equations are valid within the fluid domain. As can be seen in Fig.~\ref{mesh}, other domains are defined and specific equations are written in each of them. Around the fluid region, a realistic motionless stainless steel region is specified, with electrical conductivity equals to $0.3$ that of liquid galinstan. The region corresponding to the excitation coil is taken to be a solid conductor with electrical conductivity equals to that of copper divided by the number of turns (300). The container is contained within the bore of the electromagnet which has a complex geometry: it has been simplified here and simply modelled as a cylindrical tube made of copper, with electrical conductivity 15 times larger than galinstan. Finally, a "vacuum" regions corresponds to air and should extend to infinity, where the magnetic field obeys a harmonic equation. At the boundary of the mesh, a Robin boundary condition is applied ${\bf \nabla } a \cdot {\bf n} = - a / R$, where $R$ is the radius of the spherical extent of the mesh. That condition is correct for the dipolar component of the magnetic field only. Smaller structures are not treated rigorously but the size of the domain ($R=1.5$ for all presented results) is large enough so that small structures have been reduced significantly at radius $R$.  
At the interface between these different regions, $a$ is a continuous function. The equations are written in a weak formulation and inserted as such in a FreeFem file. 

On figure \ref{comparenum}, the numerical response is compared to the experimentally measured response for the strongest magnetic field. Parts of the experimental signal are not found in the numerical response: they may be due to structural vibrations, independent of Alfv\'en wave propagation. On fig.~\ref{pulsesadim} (right hand side), 
the same values of the Lundquist number as for the experiments are used, while the magnetic Prandtl number was taken to be $10^{-3}$: this is larger than that of Galinstan ($1.56 \, \times \, 10^{-6}$, see \cite{BuhlerMuller}), but small enough so that viscous dissipation is negligible while Hartmann boundary layers can still be resolved. 

The pulse is generated as an azimuthal electric current forced in the domain of the excitation coil. This is simply a uniform source term for the laplacian of $a$, with a temporal analytical expression that matches the experimental pulse current. 
The mesh used by FreeFem is generated by Bamg -- another free software developed by INRIA -- and can be refined in regions of strong gradients (see Fig.~\ref{mesh}). The functional space used for the calculations presented here is that of quadratic functions on each mesh element (the functions and their first derivatives are continuous functions). 
The evolution of the magnetic potential scalar and velocity streamfunction is computed using an Adams-Moulton scheme. 


\subsection{SFEM{a}NS }
\label{numpulseAdolfo}

The generation and propagation was also calculated using SFEMaNS \cite{Guer06}, a numerical code developed for cylindrical geometries using a finite element method in the meridional plane and spectral expansion in the azimuthal direction. 
We have checked that FreeFem and SFEMaNS give the same result for the same conditions. 


Let us give a brief overview of the numerical method which is used. The problem is modeled by the Full MHD equations in the eddy current approximation: induction equation, Navier-Stokes equations with Lorentz electromagnetic forces and continuity of velocity and magnetic fields. 
%
 %
%
We solve this system of equations in a heterogeneous domain composed of conducting regions of different conductivities ($\sigma_1$$>$0, $\sigma_2$$>$0,. . .) and an insulating region (vacuum, $\sigma_0$ = 0). Since the magnetic field is curl free in vacuum, it can be expressed as the gradient of a scalar potential $\phi$, as long as the insulating domain is simply connected. Enforcing continuity of $\bf B$ and $\bf \nabla\phi$ across interfaces is a significant numerical difficulty. In our Finite Element approximation, continuity is weakly imposed using an Interior Penalty Galerkin method (IPG) together with Lagrange elements. This method has been shown to be stable and convergent in \cite{Boss93,Guer06}. Since the geometry is axisymmetric, the equations are written in cylindrical coordinates and the approximate solution is expanded in Fourier series in the azimuthal direction and nodal Lagrange finite elements in the meridian plane. The time is discretized by means of a semi-implicit Backward Finite Difference method of second order (BDF2). At the boundary of the computational domain, we can impose Robin, Neumann, or Dirichlet boundary conditions.

The computations are carried out with the following discretization characteristics:
the conducting domain, composed of a finite cylinder (fluid) of 0.96 of height and 0.53 of radius and of one rectangle on the bottom represents the excitation coil from $z\in[-0.1;0]$ and $r\in[0.2;0.4]$. An additional external layer of thickness 0.02 represents the vessel. All these domains are discretized using a quasi-uniform grid of mesh size ${\Delta}$x = 1/100. The conducting region is embedded in a spherical insulating domain of radius 5 whose mesh size varies from ${\Delta}$x = 1/40 at the inner interface to ${\Delta}x$ = 0.5 at the outer boundary of the sphere. We also realize conductivity jumps compatible with the material used in the experiment.

\begin{figure}
\vspace*{-3 cm}
\hspace*{1 cm}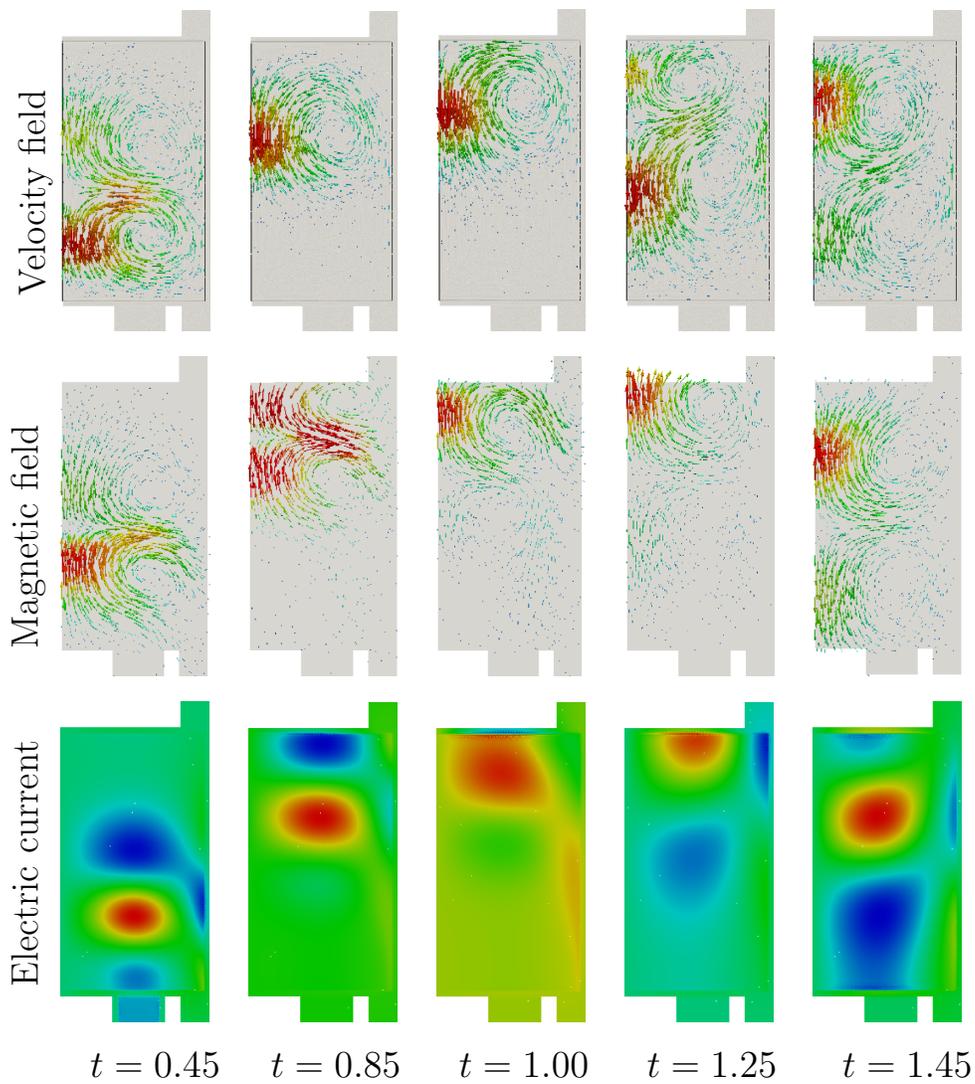
\caption{Reflexion of an Alfv\'en waves on the top end of the cylinder with SFEMaNS: meridional plane of an axisymmetrical geometry with symmetry axis on the left}
\label{reflect}
\end{figure}

On Fig.~\ref{reflect}, the reflection of an Alfv\'en wave is shown on the end of the cylinder. The Lundquist number is equal to 61.2 corresponding to an external magnetic field of 13~T and the magnetic Prandtl number is equal to 5~10$^{-3}$. We can see that the wave reaches the top of the cylinder after one dimensionless Alfv\'en time. Both counter-rotating loops of magnetic and velocity field experience severe distorsion during reflection and are then retrieved on their way back with the same configuration for the velocity field and with an opposite sign for the magnetic field (also seen on the electric current).   

\begin{figure}
\hspace*{3 cm}\includegraphics[width=10cm,keepaspectratio]{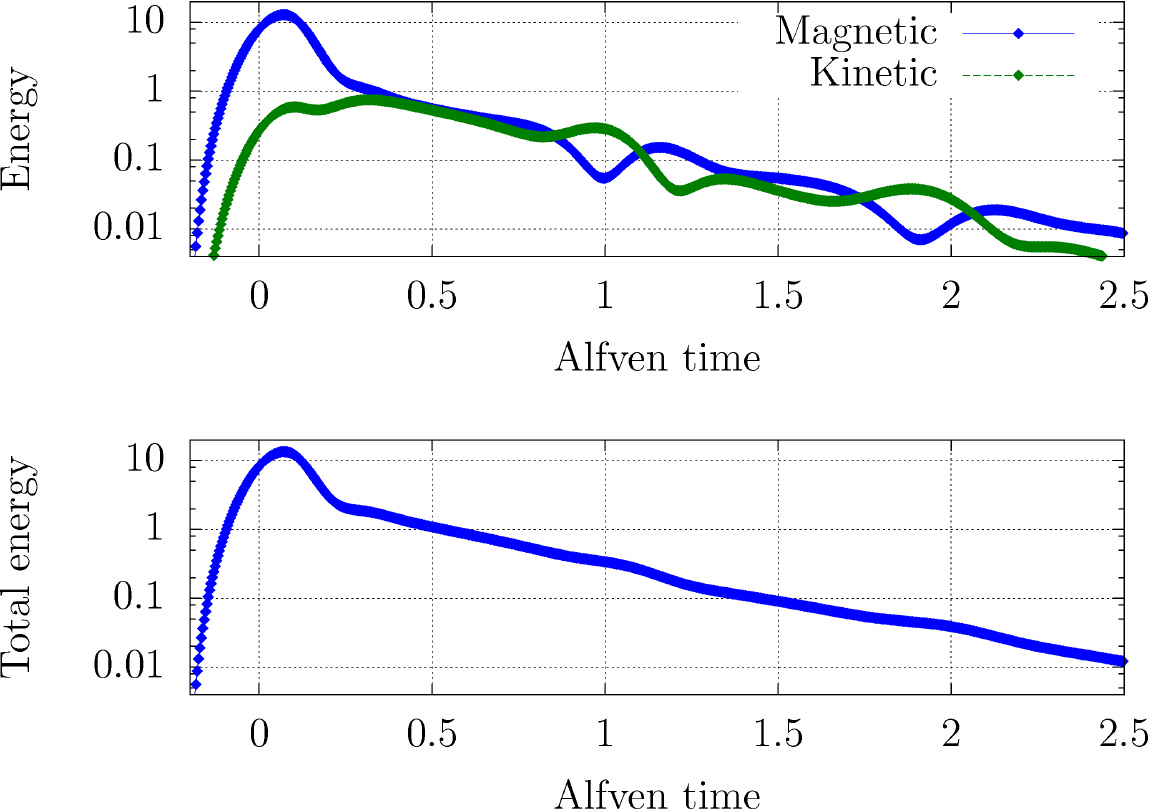}
\caption{Magnetic and kinetic energies during propagation and reflection of an Alfv\'en wave. Note that here the origin for time is taken at the peak of the electric current pulse}
\label{diss}
\end{figure}

Regarding energy, there is equipartition between kinetic and magnetic energies. In addition, there is some energy exchange during reflection between kinetic and magnetic contributions, accompanied by a net loss due to dissipation within the galinstan and within the stainless steel container (see Fig.~\ref{diss}). 

\section{Numerical simulations of harmonic response}
\label{numharmony}

It is possible to compute harmonic responses using temporal evolution calculations, but this is costly due to the necessity to go beyond the initial transient period. 
A second version of the FreeFem file has been written to determine the harmonic response. The size of the numerical problem is doubled as a real and an imaginary part are computed for $a$,  $\omega$ and $\psi$. The input current density is a pure real function. Those functions are multiplied by $e^{2 \pi i f t}$, so that time derivatives are changed into multiplications by $2 \pi i f$. This results in a purely spatial problem, providing the harmonic response at the specified frequency $f$ (see Fig.~\ref{sweepadim}, on the right hand side). 


\section{Discussion}
\label{discussion}

The experimental results presented in this paper mark some progress compared to those already published and have also some distinct features. First, a direct Alfv\'en response to a pulse is presented here for the first time: in Jameson's paper \cite{jameson}, there are only harmonic responses. The second part on the progression of Alfv\'en waves was never published, however there are some results in Jameson's PhD thesis and a glimpse of these results in the educational film by Shercliff on magnetohydrodynamics. A distinct feature of our approach is the  relative simplicity of our experiments: we have used a harmless galinstan alloy, at room temperature and any action was external to the container. We have not injected any current in the fluid, we have not inserted any search coil in the middle of the fluid region. This was possible only because we have had access to strong magnetic fields at the LNCMI. In terms of Lundquist number, we have reached approximately the same value as Jameson, with a smaller magnetic Prandtl number. 

\begin{figure}
\hspace*{-1 cm}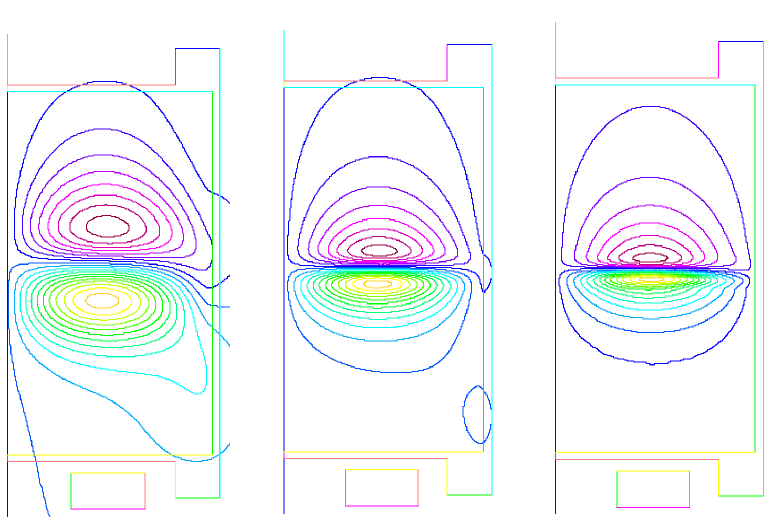 \hspace*{5mm} \includegraphics[height = 6 cm,keepaspectratio]{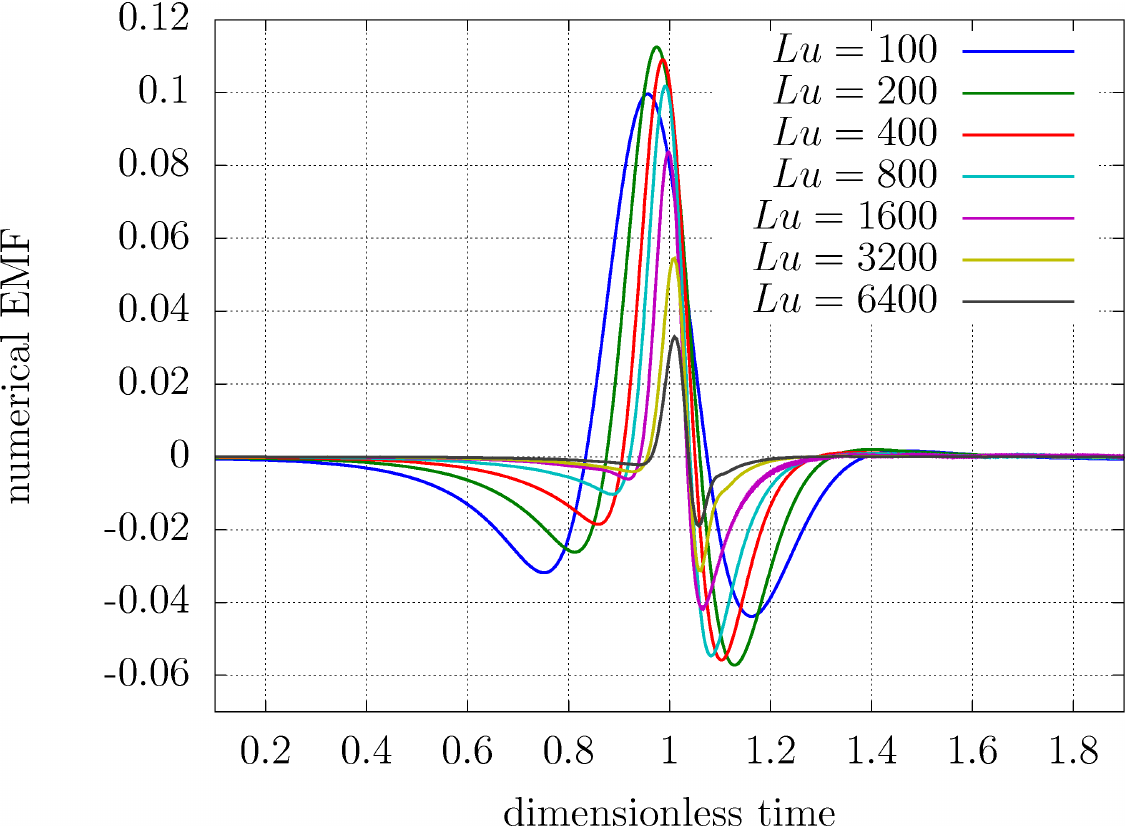}
\caption{Iso-contours of magnetic streamfunction for Lundquists number 100, 800 and 6400 at Alfv\'en time 0.5, on the left hand side. Dimensionless EMF for Lundquist numbers from 100 to 6400 measured at the top coil.}
\label{arrivals}
\end{figure}

Another distinct feature of our experiments is that velocity and magnetic fluctuations are both poloidal fields, while they were toroidal in the experiments by Lehnert, Lundquist and Jameson. This property made it possible to measure these fluctuations from outside the container. This has also consequences on the arrival time of Alfv\'en waves: the condition of continuity forces the velocity and magnetic fields to have returning components, including beyond the theoretical reach of Alfv\'en waves. This can be best seen on numerical calculations with Lundquist numbers much larger than achieved in the experiments (see Fig.~\ref{arrivals}). For those numerical runs, we have had to make a few changes: the electrical conductivity of the stainless steel container has be set to zero, the duration of the pulse has been reduced to one hundredth of Alfv\'en propagation time (making it effectively equivalent to a Dirac function). Because of these changes, the dimensionless amplitudes should not be compared with previous calculations. However, the shape of EMF response for $Lu=100$ is quite similar to the response at $Lu=61.2$ for realistic experimental conditions. For very large magnetic fields, the dimensionless amplitude becomes smaller due to diffusion effects on a short scale, {\it i.e.} the thickness of the electric current structure.  

	While vorticity and electric currents travel as thin sheets with little diffusion, magnetic and velocity fields extend on both sides on the length scale of the radius. This is directly related to the poloidal nature of velocity and magnetic fields in our configuration. The central part of the structure (vorticity and electric current sheet) reaches the top at a time very close to one at large Lundquist number. It can also be remarked that the double torus structure is generic to our configuration and valid also for arbitrarily large Lundquist numbers. 

Future experiments can make progress in two directions. First, with liquid sodium and slightly stronger magnetic fields, the Lundquist number can be increased by a factor 10. Secondly, with more energetic electrical disturbances, one may hope to excite non-linearly some resonant modes, in the same manner as inertial modes \cite{kerswell02}. 

\vspace*{2 cm}

{\bf Acknowledgments}

{\it
Thanks are due to Caroline Nore and Jacques L\'eorat for their help and discussions on the numerical simulations of Alfv\'en waves and to Dominique Jault, Henri-Claude Nataf, Nathana\"el Schaeffer and Nicolas Gillet for fruitful and extensive discussions on Alfv\'en waves. 
We thank the LNCMI laboratory for hosting our experiment and for their kind assistance: this was made possible thanks to the support of the European Commission from the 7$^{th}$ framework programme capacities ``Transnational Access'', Contract N$^\circ$ 228043-EuroMagNET II -- Integrating Activities. The project has also benefited from an internal grant from the LGIT/ISTerre laboratory.    }


\newcommand{\noopsort}[1]{} \newcommand{\printfirst}[2]{#1}
  \newcommand{\singleletter}[1]{#1} \newcommand{\switchargs}[2]{#2#1}

\end{document}